\documentclass[sigconf,screen,authorversion,nonacm]{acmart}

\pdfoutput=1

\usepackage[nolist]{acronym}
\begin{acronym}

    \acro{3DES}{Triple-DES}
    
    \acro{ACT-R}{Adaptive Control of Thought-Rational}
    \acro{AES}{Advanced Encryption Standard}
    \acro{ALU}{Arithmetic Logic Unit}
    \acro{ANOVA}{Analysis of Variance}
    \acroplural{ANOVA}[ANOVAs]{Analyses of Variance}
    \acro{API}{Application Programming Interface}
    \acro{AOI}{Area of Interest}
    \acroplural{AOI}[AOIs]{Areas of Interest}
    \acro{ARX}{Addition Rotation XOR}
    \acro{ATPG}{Automatic Test Pattern Generation}
    \acro{ASIC}{Application Specific Integrated Circuit}
    \acro{ASIP}{Application Specific Instruction-Set Processor}
    \acro{AS}{Active Serial}
    
    \acro{BDD}{Binary Decision Diagram}
    \acro{BGL}{Boost Graph Library}
    \acro{BKA}{German Federal Criminal Police Office}
    \acro{BNF}{Backus-Naur Form}
    \acro{BRAM}{Block-Ram}
    
    \acro{CBC}{Cipher Block Chaining}
    \acro{CFB}{Cipher Feedback Mode}
    \acro{CFG}{Control Flow Graph}
    \acro{CLB}{Configurable Logic Block}
    \acro{CLI}{Command Line Interface}
    \acro{CMOS}{Complementary Metal-Oxide-Semiconductor}
    \acro{COFF}{Common Object File Format}
    \acro{CPA}{Correlation Power Analysis}
    \acro{CPU}{Central Processing Unit}
    \acro{CRC}{Cyclic Redundancy Check}
    \acro{CTA}{Concurrent Think Aloud}
    \acro{CTR}{Counter}
    
    \acro{DC}{Direct Current}
    \acro{DES}{Data Encryption Standard}
    \acro{DFA}{Differential Frequency Analysis}
    \acro{DFT}{Discrete Fourier Transform}
    \acro{DIP}{Distinguishing Input Pattern}
    \acro{DLL}{Dynamic Link Library}
    \acro{DMA}{Direct Memory Access}
    \acro{DNF}{Disjunctive Normal Form}
    \acro{DPA}{Differential Power Analysis}
    \acro{DSO}{Digital Storage Oscilloscope}
    \acro{DSP}{Digital Signal Processing}
    \acro{DUT}{Design Under Test}
    
    \acro{ECB}{Electronic Code Book}
    \acro{ECC}{Elliptic Curve Cryptography}
    \acro{EEPROM}{Electrically Erasable Programmable Read-only Memory}
    \acro{EMA}{Electromagnetic Emanation}
    \acro{EM}{electro-magnetic}
    \acro{ET}[eye tracking]{eye tracking} 
    \acro{EU}{European Union}
    
    \acro{FFT}{Fast Fourier Transformation}
    \acro{FF}{Flip Flop}
    \acro{FI}{Fault Injection}
    \acro{FIR}{Finite Impulse Response}
    \acro{FPGA}{Field Programmable Gate Array}
    \acro{FSM}{Finite State Machine}
    
    \acro{GT}{Grounded Theory}
    \acro{GUI}{Graphical User Interface}
    
    \acro{HCI}{Human Computer Interaction}
    \acro{HDL}{Hardware Description Language}
    \acro{HD}{Hamming Distance}
    \acro{HF}{High Frequency}
	\acro{HRE}{Hardware Reverse Engineering}
    \acro{HSM}{Hardware Security Module}
    \acro{HW}{Hamming Weight}
    
    \acro{IC}{Integrated Circuit}
    \acro{IO}[I/O]{Input/Output}
    \acro{IOB}{Input Output Block}
    \acro{IoT}{Internet of Things}
    \acro{IRB}{Institutional Review Board}
    \acro{IP}{Intellectual Property}
    \acro{IRR}{Inter-Rater Reliability}
    \acro{ISA}{Instruction Set Architecture}
    \acro{ISCED}{International Standard Classification of Education}
    \acro{IV}{Initialization Vector}
    
    \acro{JTAG}{Joint Test Action Group}
    
    \acro{KAT}{Known Answer Test}
    
    \acro{LFSR}{Linear Feedback Shift Register}
    \acro{LSB}{Least Significant Bit}
    \acro{LUT}{Look-up table}
    
    \acro{MAC}{Message Authentication Code}
    \acro{MAD}{Median Absolute Deviation}
    \acro{MIPS}{Microprocessor without Interlocked Pipeline Stages}
    \acro{MMIO}{Memory Mapped \acl{IO}}
    \acro{MSB}{Most Significant Bit}
    
    \acro{NASA}{National Aeronautics and Space Administration}
    \acro{NSA}{National Security Agency}
    \acro{NVM}{Non-Volatile Memory}
    
    \acro{OEM}{Original Equipment Manufacturer}
    \acro{OFB}{Output Feedback Mode}
    \acro{OISC}{One Instruction Set Computer}
    \acro{ORAM}{Oblivious Random Access Memory}
    \acro{OS}{Operating System}
    
    \acro{PAR}{Place-and-Route}
    \acro{PCB}{Printed Circuit Board}
    \acro{PC}{Personal Computer}
    \acro{PS}{Processing Speed}
	\acro{PR}{Perceptual Reasoning}
    \acro{PUF}{Physically Unclonable Function}

    \acro{QA}{Quality Assurance}
    
    \acro{RISC}{Reduced Instruction Set Computer}
    \acro{RNG}{Random Number Generator}
    \acro{ROBDD}{reduced ordered binary decision diagram}
    \acro{ROM}{Read-Only Memory}
    \acro{ROP}{Return-oriented Programming}
    \acro{RTA}{Retrospective Think Aloud}
    \acro{RTL}{Register-Transfer Level}
    \acro{RUB}{Ruhr University Bochum}
    
    \acro{SAT}{Boolean satisfiability}
    \acro{SEMOBS}{Self-Modifying Bitstreams}
    \acro{SCA}{Side-Channel Analysis}
    \acro{SHA}{Secure Hash Algorithm}
    \acro{SNR}{Signal-to-Noise Ratio}
    \acro{SPA}{Simple Power Analysis}
    \acro{SPI}{Serial Peripheral Interface Bus}
    \acro{SRAM}{Static Random Access Memory}
    \acro{SRE}{Software Reverse Engineering}

    \acro{TA}{Think Aloud}
	
	\acro{VLSI}{Very-Large-Scale Integration}
	
	\acro{WAIS-IV}{Wechsler Adult Intelligence Scale}
	\acro{WM}{Working Memory}
    
    \acro{TRNG}{True Random Number Generator}
    
    \acro{UART}{Universal Asynchronous Receiver Transmitter}
    \acro{UHF}{Ultra-High Frequency}
    \acro{UI}{User Interface}
    
    \acro{vFPGA}{virtual FPGA}
	\acro{VC}{Verbal Comprehension}
    
    \acro{WISC}{Writeable Instruction Set Computer}
    
    \acro{XDL}{Xilinx Description Language}
    \acro{XTS}{XEX-based Tweaked-codebook with ciphertext Stealing}

    \acro{ZITiS}{German Central Office for Information Technology in the Security Sector}
    
\end{acronym}

\usepackage{hyperref}
\usepackage{booktabs}
\usepackage{csquotes}
\usepackage{enumerate}
\usepackage{rotating}
\usepackage{siunitx}
\usepackage{pdfpages}
\usepackage[inline]{enumitem}
\usepackage{xspace}
\usepackage{tikz}
\usepackage{xcolor}
\usepackage{listings}
\usepackage{pgfkeys}
\usepackage{footmisc}

\usetikzlibrary{positioning}
\usetikzlibrary{backgrounds}

\definecolor{colKeys}{rgb}{0,0,1}
\definecolor{colIdentifier}{rgb}{0,0,0}
\definecolor{colComments}{rgb}{1,0,0}
\definecolor{colString}{rgb}{0,0.5,0}

\newcommand{\coursekey}[1]{\textbf{\pgfkeysvalueof{/courses/#1/key}}}

\newcommand{\coursecite}[1]{\cite{\pgfkeysvalueof{/courses/#1/ref}}}

\input{courses.pgfkeys}

\newenvironment{inumerate}{
    \begin{enumerate*}[itemjoin={{, }}, itemjoin*={{, and }}, after={{.}}]
}{
    \end{enumerate*}
}

\newcommand{\ie}{i.\,e.}
\newcommand{\eg}{e.\,g.}

\newcommand{\etal}{et~al.\@\xspace}

\newcommand*\fullcirc[1][0.8ex]{\tikz\fill (0,0) circle (#1);} 

\newcommand*\rot{\rotatebox{90}}

\DefineFNsymbolsTM{nodagger}{
  \textasteriskcentered *
  \textdaggerdbl \ddagger
  \textsection   \mathsection
  \textbardbl    \|%
  \textparagraph \mathparagraph
}%
\setfnsymbol{nodagger}

\copyrightyear{2025}
\acmYear{2025}
\setcopyright{rightsretained}
\acmConference[SIGCSE TS 2025]{Proceedings of the 56th ACM Technical Symposium on Computer Science Education V. 1}{February 26-March 1, 2025}{Pittsburgh, PA, USA}
\acmBooktitle{Proceedings of the 56th ACM Technical Symposium on Computer Science Education V. 1 (SIGCSE TS 2025), February 26-March 1, 2025, Pittsburgh, PA, USA}
\acmDOI{10.1145/3641554.3701797}
\acmISBN{979-8-4007-0531-1/25/02}

\begin{document}

\title{An Evidence-Based Curriculum Initiative for Hardware Reverse Engineering Education}

\author{René Walendy} 
\orcid{0000-0002-5378-3833}
\email{rene.walendy@rub.de}
\affiliation{%
    \institution{Ruhr University Bochum}
    \city{Bochum}
    \country{Germany}
}
\additionalaffiliation{%
    \institution{Max Planck Institute for Security and Privacy}
    \city{Bochum}
    \country{Germany}
}

\author{Markus Weber} 
\orcid{0000-0001-7775-807X}
\email{markus.weber3@rub.de}
\affiliation{%
    \institution{Ruhr University Bochum}
    \city{Bochum}
    \country{Germany}
}

\author{Steffen Becker} 
\orcid{0000-0001-7526-5597}
\email{steffen.becker@rub.de}
\affiliation{%
    \institution{Ruhr University Bochum}
    \city{Bochum}
    \country{Germany}
}
\authornotemark[1] 

\author{Christof Paar} 
\orcid{0000-0001-8681-2277}
\email{christof.paar@mpi-sp.org}
\affiliation{%
    \institution{Max Planck Institute for Security and Privacy}
    \city{Bochum}
    \country{Germany}
}

\author{Nikol Rummel} 
\orcid{0000-0002-3187-5534}
\email{nikol.rummel@rub.de}
\affiliation{%
    \institution{Ruhr University Bochum}
    \city{Bochum}
    \country{Germany}
}
\additionalaffiliation{%
    \institution{Center for Advanced Internet Studies (CAIS)}
    \city{Bochum}
    \country{Germany}
}

\renewcommand{\shortauthors}{René Walendy, Markus Weber, Steffen Becker, Christof Paar, and Nikol Rummel}

\begin{abstract}
    The increasing importance of supply chain security for digital devices---from consumer electronics to critical infrastructure---has created
    a high demand for skilled cybersecurity experts.
    These experts use \acf{HRE} as a crucial technique to ensure trust in digital semiconductors.
    Recently, the US and EU have provided substantial funding to educate this cybersecurity-ready semiconductor workforce, but success depends on the widespread availability of academic training programs.
    In this paper, we investigate the current state of education in hardware security and \ac{HRE} to identify efficient approaches for establishing effective \ac{HRE} training programs.
    Through a systematic literature review, we uncover 13 relevant courses, including eight with accompanying academic publications.
    We identify common topics, threat models, key pedagogical features, and course evaluation methods.
    We find that most hardware security courses do not prioritize \ac{HRE}, making \ac{HRE} training scarce.
    While the predominant course structure of lectures paired with hands-on projects appears to be largely effective, we observe a lack of standardized evaluation methods and limited reliability of student self-assessment surveys.
    Our results suggest several possible improvements to \ac{HRE} education and offer recommendations for developing new training courses.
    We advocate for the integration of \ac{HRE} education into curriculum guidelines to meet the growing societal and industry demand for \ac{HRE} experts.
\end{abstract}





\maketitle

\section{Introduction} \label{chap:intro}

Digital microelectronics in the form of \acfp{IC} have become ubiquitous in today's digital society.
To be able to trust the various systems we interact with every day, it is essential that not only their software, but even more so the underlying hardware, is trustworthy.
Assuring trust in hardware, however, is a major challenge:
Most \acp{IC} are not manufactured domestically, making the global semiconductor supply chain a lucrative target for insider and outsider attackers seeking to compromise hardware reliability and integrity~\cite{bhunia2014hardware}, or gain unfair economic advantage~\cite{guin2014counterfeit}.
Such hardware vulnerabilities and manipulations pose a serious threat to the privacy and security of both individuals and entire industries.

The US and EU have recently begun to address this concern through massive regulatory efforts~\cite{rep.ryan20224346, eu2023regulation}, calling for urgent measures and providing substantial funding to ramp up educational programs for a cybersecurity-ready semiconductor workforce.
The EU and US semiconductor industries require highly skilled hardware security experts to achieve two major goals:\begin{inumerate}%
    \item in the long term, to increase the domestic production of trustworthy semiconductors
    \item in the medium term, to validate designs and parts procured through the global \ac{IC} supply chain%
\end{inumerate}

\acf{HRE} is an important technique particularly for the latter goal.
As it can be used to both protect and attack \acp{IC}~\cite{quadir2017survey,wallat2017look}, making \ac{HRE} knowledge more available to defenders is important in the arms race against resourceful and often state-sponsored attackers.
Despite strong industry demand, there exist few dedicated hardware security courses at academic institutions worldwide~\cite{cabaj2018cybersecurity}.
Many of the major courses in the field focus primarily on side-channel protection measures.
Thus, we see a need for courses more comprehensively covering the \ac{HRE} skills required for supply chain validation.

\paragraph{Survey on \texorpdfstring{\acs{HRE}}{HRE} Education}
\label{par:introduction:survey}
To empirically assess the current demand for educational programs in \ac{HRE}, we conducted anonymous surveys at two leading events in the field: the academically focused HARRIS workshop and the industry-oriented Hardwear.io conference.
We collected 68 responses with participant experience ranging from none (3 participants) over up to three years (30), 4-6 years (10), 7-9 years (2), to at least ten years (18).
Five participants did not report years of experience.
Participants rated the importance of four learning environments for acquiring expertise on a Likert scale (1=not important, 5=very important).
Independent learning was rated as most important ($mean = 4.07, SD = 1.01$), followed by on-the-job learning ($mean = 3.79, SD = 1.45$), workshops ($mean = 3.38, SD = 1.12$), and academia ($mean = 2.91, SD = 1.28$).
Participants also rated four factors affecting the difficulty of entering the \ac{HRE} field (1=not at all, 5=extremely).
The greatest hurdle was the unavailability of mentors ($mean = 3.51, SD = 1.2)$, followed by the lack of affordable tooling ($mean = 3.19, SD = 1.33$) and educational programs ($mean = 3.17, SD = 1.3$).

It is therefore worthwhile for academic institutions to broaden their cybersecurity curricula.
Specifically, incorporating dedicated \ac{HRE} training into more programs will better prepare the next generation of hardware security professionals to meet industry demand.
To this end, instructors would benefit from more thorough research into the optimal structure for academic \ac{HRE} courses, expanding on the presently few recommendations in prominent curriculum guidelines~\cite{jtfcc2018cybersecurity}.
Identifying limiting factors and ways to overcome them can further support the setup of new \ac{HRE} courses.

\paragraph{Research Questions}
\label{par:introduction:rqs}
With the overarching goal of identifying efficient approaches for establishing \ac{HRE} training programs to meet the demand for a cybersecurity-ready domestic semiconductor workforce, we answer the following research questions:
\begin{description}
    \item[RQ1a] What are the main contents of existing hardware security courses? Do they comprise \ac{HRE}?
    \item[RQ1b] What are the typical pedagogical features of existing hardware security courses?
    \item[RQ2a] What evidence is there on their effectiveness in teaching knowledge of and skills for \ac{HRE}?
    \item[RQ2b] Are the typical methods for evaluating course effectiveness adequate?
\end{description}

\section{Background}
\label{chap:background}

\subsection{The Need for \texorpdfstring{\acs{HRE}}{HRE} Education}
\label{sec:background:hwsec}

Computer security research has long focused on \textit{software} security, treating the underlying hardware that executes it as an inherently trustworthy environment.
With the discovery of hardware-level vulnerabilities such as Spectre~\cite{kocher2020spectre} in 2019, cybersecurity researchers more and more have come to realize: Any (software) system can only ever be as secure as the hardware running it~\cite{bloom2012hardware}.

\subsubsection{The US CHIPS and Science Act and the European Chips Act}
\label{sec:background:hwsec:chipsact}

As the problem of secure hardware at the \ac{IC} level arises during production, it cannot be solved by widespread end user education.
Instead, highly skilled experts trained in hardware verification are urgently needed to thwart attacks on the complex \ac{IC} supply chains and maintain the digital sovereignty of both the EU and the US.

Both governments have recently launched large-scale initiatives to this end, namely the European Chips Act of 2023~\cite{eu2023regulation} and the US CHIPS Act of 2022~\cite{rep.ryan20224346}.
Either aims to secure the semiconductor supply through massive investment in domestic---hence independent---manufacturing and testing capabilities, with a projected workforce growth of more than one million additional workers until 2030~\cite{weisz2022global}.

\subsubsection{Hardware Reverse Engineering for Security Assurance}
\label{sec:background:hwsec:assurance}
A primary technique in hardware verification is \acf{HRE}, the \enquote{act of creating a set of specifications for a piece of hardware by someone other than the original designers, primarily based upon analyzing and dimensioning a specimen.}~\cite{rekoff1985reverse}

First, it supports large-scale hardware verification~\cite{tehranipoor2010survey}, which often works by comparing the product to an exhaustively validated \textit{golden model}.
\ac{HRE} can prove this model to be genuine and free of malicious manipulations~\cite{bhunia2014hardware,puschner2023red}, commonly called hardware Trojans~\cite{becker2019hardware}. 
Second, \ac{HRE} allows hardware security analysts to discover new vulnerabilities and previously unseen or particularly subtle types of tampering that may elude other verification techniques aimed at detecting signatures of well-known attacks~\cite{ludwig2021vital,lippmann2023comprehensive}.
Thus, \ac{HRE} helps to gain an advantage in the hardware security arms race.

\subsection{Curriculum Guidelines for \texorpdfstring{\acs{HRE}}{HRE} Education}
\label{sec:background:education}

While both the European Chips Act and US CHIPS Act provide extensive funding and frameworks for educational programs~\cite[Sec. 10318]{rep.ryan20224346}, \cite[Operational Objective 4]{eu2023regulation}, various scholars have warned that the current state of security education is inadequate to meet the industry demand for a cybersecurity-ready workforce~\cite{cabaj2018cybersecurity, goupil2022understanding, blazic2021cybersecurity}.
This systematic shortage appears to be even more pronounced in the \ac{HRE} subfield~\cite{cohen2018need}, which we argue is partly caused by a lack of actionable curriculum guidelines.
For instance, the just-released iteration of the joint ACM and IEEE curriculum guideline for computer science~\cite{jtfcc2023computerfinal} makes no provision for \ac{HRE}-specific content---despite mentioning the term \enquote{security} more than 400, and \enquote{hardware} more than 100 times.
The 2017 guideline for post-secondary cybersecurity education~\cite{jtfcc2018cybersecurity} does call for the inclusion of both \ac{HRE} techniques and countermeasures as well as supply chain security in the \enquote{Component Security} knowledge area.
However, it does not provide any suggestions for course packaging.
Other guidelines still consider hardware assurance as an optional knowledge area, despite pressing concerns from the US Department of Defense~\cite{cohen2018need}.
In this paper we analyze to which extent \ac{HRE} education has proliferated and how it could benefit from specific curriculum guidance.

\section{Methods}
\label{chap:methods}
We conducted a systematic literature review\footnote{Detailed data is available on the Open Science Framework: \url{https://osf.io/dt8ne/}} to identify and synthesize a corpus of academic courses with \ac{HRE} content, and their accompanying publications that might shed light on the pedagogical considerations underlying the courses.
We chose this approach over a web search for course listings with the goal of replicability in mind, rather than out of any consideration of course quality.

\subsection{Sampling Procedure}
\label{sec:methods:sampling}
We queried the electronic literature database Semantic Scholar using its Academic Graph bulk search interface~\cite{kinney2023semantic} on May 1, 2024.
We chose Semantic Scholar as the single search provider because it aggregates publications from numerous literature databases under a single, well-defined search interface.
It covers major publishers in computer science and computer science education, such as the ACM Digital Library, IEEE Xplore, DBLP, and Wiley Online Library.
It also includes gray literature through the arXiv preprint library.

In our initial corpus, we included any result that matched the following search string in its title or abstract, with no limit on publication date:
\begin{lstlisting}
    ("hardware reverse engineering" | "hardware security") + (course | teaching | education)
\end{lstlisting}

We decided to include the broader term \enquote{hardware security} as an alternative to \enquote{hardware reverse engineering} in order to discover related courses that may still offer limited \ac{HRE} content.

\subsection{Selection Process}
\label{sec:methods:selection}

We manually screened the 66 results in two sifts, each under the following inclusion criteria:
\begin{itemize}
    \item Full research papers written in English
    \item Works presenting full-semester, for-credit academic courses
\end{itemize}
We further applied the following exclusion criteria:
\begin{itemize}
    \item Results not pertaining to an individual academic research publication, such as journal editorials
    \item Courses not primarily about hardware security or \ac{HRE}
\end{itemize}

For the first sift, we considered title and abstract only, leaving 20 publications.
For the second sift, we read these papers in full and re-evaluated the above criteria, resulting in a collection of nine publications.
To ensure broad coverage, we repeated our search using a second well-known database, Web of Science, on November 4, 2024.
This second search yielded no additional results after the first sift.

\subsection{Collection of Course Materials}
\label{sec:methods:materials}

For all the courses described in the publications, we obtained publicly available materials through the teaching websites of the respective institutions and the personal websites of the instructors involved.
Where material was no longer publicly available, we used the {Internet Archive\footnote{\url{https://web.archive.org/}}} to obtain historical versions where possible.

We made the best effort to add related courses \cite{farahmandi2019teaching,forteNNteaching,tehranipoor2012hardware, tehranipoor2019trustable} taught at the same institution to our corpus, even when they were not covered by a scholarly publication. 
This approach allows us to better gauge the breadth of \ac{HRE} and hardware security education at a given institution.

During collection of courses, we made one notable exception to our strategy:
Through discussions with instructors in the field, we became aware of an \ac{HRE} course developed at Rensselaer Polytechnic Institute.
While there is no formal publication on this course, most of the course materials are freely available online.
Since data on \ac{HRE} courses is scarce, we decided to still include this course, resulting in a corpus of 13 academic courses from nine institutions.

\subsection{Analysis}
\label{sec:methods:analysis}
We identified the main topics and structure of each course, as well as the materials used in the respective lectures and assignments, from syllabi and lecture slides to the extent available, and from the corresponding publications otherwise.
Timelines were extracted either from the publications or from historical versions of the course websites.
From this data, we distilled a set of major threat models considered in the courses.
We also identified all means of didactic evaluation described in the corresponding publications, as well as the key findings of these evaluations.

\section{Results}
\label{chap:results}

In this section, we present the findings from our systematic literature review, including historical developments and the key topics covered in \textit{hardware security} and \textit{\acl{HRE}} courses.
We derive the distinction between these types of courses and analyze the resource requirements for establishing similar courses at other institutions.

\subsection{Overview of Discovered Courses}
\label{sec:results:history}

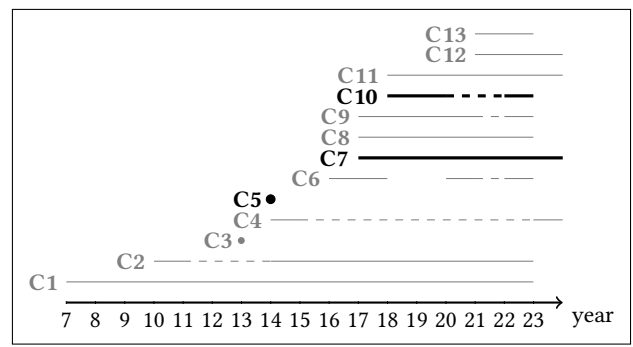
\begin{figure}[b]
    \centering
    \fbox{\begin{tikzpicture}[scale=.55]
        \draw[thick,->] (0,0) -- (12,0) node[anchor=north west] {year};
        \foreach \x [count=\xi] in {7,...,23}
            \draw ({(\xi-1)/17*12},1pt) -- ({(\xi-1)/17*12},-1pt) node[anchor=north] {$\x$};

        \draw[gray] (0,0.5) node[anchor=east] {\coursekey{embedded}} -- (11.29,0.5);
        \draw[gray] (2.12,1) node[anchor=east] {\coursekey{introtrust}} -- (2.82,1); \draw[dashed,gray] (2.82,1) -- (4.94,1); \draw[gray] (4.94,1) -- (11.29,1);
        \filldraw[gray] (4.23,1.5) circle (2pt) node[anchor=east] {\coursekey{trustable}} ;
        \draw[gray] (4.94,2) node[anchor=east] {\coursekey{halak}} -- (5.65,2); \draw[dashed,gray] (5.65,2) -- (11.29,2); \draw[gray] (11.29,2) -- (12,2);
        \filldraw (4.94,2.5) circle (3pt) node[anchor=east] {\coursekey{zonenberg}} ;
        \draw[gray] (6.35,3) node[anchor=east] {\coursekey{advtrust}} -- (7.76,3); \draw[gray] (9.18,3) -- (9.88,3); \draw[dashed,gray] (9.88,3) -- (10.59,3); \draw[gray] (10.59,3) -- (11.29,3);
        \draw[very thick] (7.06,3.5) node[anchor=east] {\coursekey{rub}} -- (12,3.5);
        \draw[gray] (7.06,4) node[anchor=east] {\coursekey{handson}} -- (11.29,4);
        \draw[gray] (7.06,4.5) node[anchor=east] {\coursekey{carpenter}} -- (9.88,4.5); \draw[dashed,gray] (9.88,4.5) -- (10.59,4.5); \draw[gray] (10.59,4.5) -- (11.29,4.5);
        \draw[very thick] (7.76,5) node[anchor=east] {\coursekey{phiks}} -- (9.18,5); \draw[dashed, very thick] (9.18,5) -- (10.59,5); \draw[very thick] (10.59,5) -- (11.29,5);
        \draw[gray] (7.76,5.5) node[anchor=east] {\coursekey{crypteng}} -- (12,5.5);
        \draw[gray] (9.88,6) node[anchor=east] {\coursekey{cad}} -- (12,6);
        \draw[gray] (9.88,6.5) node[anchor=east] {\coursekey{practical}} -- (11.29,6.5);
    \end{tikzpicture}}
    \caption{Timeline of university courses focused on hardware security. Dots mark courses with only a single iteration; dashed lines signify unclear continuity of the course. Courses with a strong focus on \ac{HRE} are highlighted.}
    \Description[Years taught, indicated for all 13 courses]{One timeline for each of the 13 courses, indicating first and last year taught, and potential gaps. Most courses start in the 2017 academic year or later. The first course started in 2007. Three courses have a strong HRE focus.}
    \label{fig:results:timeline}
\end{figure}

\autoref{fig:results:timeline} shows a year-by-year summary of the 13 hardware security and \ac{HRE} course timelines.
The first offer, \coursekey{embedded}\footnote{For brevity, we assigned an identifier \textbf{Cx} to each course. See \autoref{tab:results:coverage:matrix} for references to each course.}, was established at Rice University in 2007.
It is still running with 14 iterations taught including the 2023/24 term, although it has since moved with its instructor to the University of California, San Diego.
This makes it the longest running course in our corpus.
A second program, \coursekey{introtrust}, was established at the University of Connecticut in 2010.
In 2012, its instructor published one of the first textbooks on hardware security~\cite{tehranipoor2012introduction}, which has been adopted by several other courses including \coursekey{carpenter} and \coursekey{practical}.
Similar courses soon after started in Connecticut (\coursekey{trustable}) and at the University of Florida (\coursekey{advtrust}, \coursekey{handson}, \coursekey{cad}), where the original instructor of \coursekey{introtrust} moved in 2016.
With five courses, University of Florida is the most represented institution in our corpus.

While the above courses cover a wide range of topics in hardware security, the first dedicated course on \ac{HRE}, \coursekey{zonenberg}, was introduced at Rensselaer Polytechnic Institute in 2014.
In 2017, a dedicated course on \textit{netlist} (\ie, logic circuit) reverse engineering, \coursekey{rub}, was launched at Ruhr University Bochum.
Finally, the third dedicated \ac{HRE} course, \coursekey{phiks}, was added to the Florida curriculum in 2018.

Other general hardware security courses emerged at independent institutions between 2014 and 2021, namely \coursekey{halak} at Southampton University, \coursekey{carpenter} at Wentworth Institute of Technology, \coursekey{crypteng} at North Carolina State University, and the most recent addition, \coursekey{practical} at the University of South Florida.

\subsection{Course Contents: Four common themes while \texorpdfstring{\acs{HRE}}{HRE} is rare}
\label{sec:results:coverage}

\bgroup
\setlength\tabcolsep{2pt}
\def\arraystretch{.9}
\begin{table}[b]
    \centering
    \caption{Matrix of common course contents. The three courses with strong \ac{HRE} focus are highlighted on the left.}
    \label{tab:results:coverage:matrix}
    \begin{tabular}{ccc*{2}{>{\centering}b{19pt}}*{14}{c}}
                                           &&& \multicolumn{2}{c}{{\parbox{44pt}{\textbf{Cryptology \& Security}}}} & \phantom{x} & \multicolumn{2}{c}{\textbf{\acs{VLSI}}} & \phantom{x} & \multicolumn{5}{c}{\parbox{45pt}{\textbf{Attacks \& Defenses}}}                  & \phantom{x} & \multicolumn{4}{c}{\textbf{\acs{HRE}}}                         \\
                                                      \cline{4-5}                                                              \cline{7-8}                                \cline{10-14}                                                                        \cline{16-19}
        \rot{\textbf{Course}} & \rot{Reference} & \phantom{x} & \rot{Essentials} & \rot{Advanced}                                     && \rot{IC Design} & \rot{Verification~}   && \rot{\acs{SCA}} & \rot{Trojans} & \rot{\acsp{PUF}} & \rot{Piracy} & \rot{Faults} && \rot{Invasive} & \rot{\acs{PCB}} & \rot{Image} & \rot{Netlist} \\
        \cline{1-2}                                   \cline{4-5}                                                              \cline{7-8}                                \cline{10-14}                                                                        \cline{16-19}
        \color{gray}\coursekey{embedded}  & \coursecite{embedded}   && -                & -                                                  && -               & -                     && \fullcirc       & \fullcirc     & \fullcirc        & \fullcirc    & -            && -              & -               & -          & -\rule{0pt}{1em}\\
        \color{gray}\coursekey{introtrust}& \coursecite{introtrust} && \fullcirc        & -                                                  && \fullcirc       & \fullcirc             && \fullcirc       & \fullcirc     & \fullcirc        & \fullcirc    & \fullcirc    && \fullcirc      & -               & \fullcirc  & -              \\
        \color{gray}\coursekey{trustable} & \coursecite{trustable}  && \fullcirc        & -                                                  && \fullcirc       & -                     && \fullcirc       & \fullcirc     & -                & -            & -            && -              & -               & -          & -              \\
        \color{gray}\coursekey{halak}     & \coursecite{halak}      && \fullcirc        & -                                                  && \fullcirc       & \fullcirc             && \fullcirc       & \fullcirc     & \fullcirc        & \fullcirc    & -            && -              & -               & -          & -              \\
                    \coursekey{zonenberg} & \coursecite{zonenberg}  && -                & -                                                  && \fullcirc       & -                     && \fullcirc       & -             & -                & -            & \fullcirc    && \fullcirc      & \fullcirc       & \fullcirc  & -              \\
        \color{gray}\coursekey{advtrust}  & \coursecite{advtrust}   && \fullcirc        & \fullcirc                                          && -               & -                     && \fullcirc       & -             & \fullcirc        & \fullcirc    & \fullcirc    && \fullcirc      & -               & -          & -              \\
                    \coursekey{rub}       & \coursecite{rub}        && \fullcirc        & -                                                  && \fullcirc       & -                     && -               & \fullcirc     & -                & \fullcirc    & -            && \fullcirc      & \fullcirc       & \fullcirc  & \fullcirc      \\
        \color{gray}\coursekey{handson}   & \coursecite{handson}    && \fullcirc        & -                                                  && -               & -                     && \fullcirc       & \fullcirc     & \fullcirc        & -            & \fullcirc    && -              & \fullcirc       & -          & -              \\
        \color{gray}\coursekey{carpenter} & \coursecite{carpenter}  && \fullcirc        & -                                                  && \fullcirc       & -                     && \fullcirc       & \fullcirc     & \fullcirc        & \fullcirc    & -            && -              & -               & -          & \fullcirc      \\
                    \coursekey{phiks}     & \coursecite{phiks}      && -                & -                                                  && -               & \fullcirc             && -               & \fullcirc     & -                & \fullcirc    & \fullcirc    && \fullcirc      & \fullcirc       & \fullcirc  & -              \\
        \color{gray}\coursekey{crypteng}  & \coursecite{crypteng}   && \fullcirc        & \fullcirc                                          && -               & -                     && \fullcirc       & \fullcirc     & \fullcirc        & -            & \fullcirc    && -              & -               & -          & -              \\
        \color{gray}\coursekey{cad}       & \coursecite{cad}        && -                & -                                                  && \fullcirc       & \fullcirc             && \fullcirc       & \fullcirc     & -                & \fullcirc    & \fullcirc    && -              & -               & -          & -              \\
        \color{gray}\coursekey{practical} & \coursecite{practical}  && -                & \fullcirc                                          && -               & -                     && \fullcirc       & \fullcirc     & \fullcirc        & -            & -            && -              & -               & -          & -              
    \end{tabular}
    \smallskip
    \raggedright\footnotesize
    \acs{VLSI}: \acl{VLSI}. \acs{SCA}: \acl{SCA}. \\ \acsp{PUF}: \aclp{PUF}. \acs{PCB}: \acl{PCB}.
    \Description[A matrix mapping courses to contents]{A matrix with one row for each course and one column for each topic. Dots indicate that a specific course contains a topic, dashes indicate the opposite. The first two columns indicate the course abbreviation and the reference from the literature search. The remaining 13 columns contain the topics. The topics are grouped into the four categories from Section 4.2.}
\end{table}
\egroup

Here we present an overview of the four major topics commonly covered by the courses in our corpus.
\autoref{tab:results:coverage:matrix} shows the detailed topics for each course\footnote{The full dataset is available on the Open Science Framework: \url{https://osf.io/dt8ne/}}.
\begin{description}
    \item[Cryptology \& IT Security]
        covers introductory course content on essential security concepts, symmetric and public-key cryptographic algorithms, ethical considerations, and various advanced topics in modern cryptography.
    \item[\acf{VLSI}]
        introduces \ac{IC} design, optimization, and manufacturing, as well as pre- and post-silicon design verification and failure analysis.
    \item[Attacks \& Defenses]
        presented across all courses were numerous, with the five most commonly observed topics being \acf{SCA} and how to avoid these side channels during design, hardware Trojans and their detection, implementing \acfp{PUF}, \ac{IC} piracy and how to detect such counterfeits, and the use and detection of fault attacks.
    \item[\acf{HRE}]
        covers the various steps in the process of recovering the design specifications from a physical device.
        At the physical level, this includes invasive attacks such as destructively imaging or probing an \ac{IC}, and reverse engineering \acfp{PCB}.
        At the digital level, \ac{IC} images can be further reverse engineered using semi-automated techniques, including machine vision, to recover the circuit diagram.
        Finally, the recovered netlist can be analyzed to reconstruct the high-level specification.
\end{description}

We find that 
\ac{HRE} content is covered by only seven of the thirteen courses.
For all except the three dedicated \ac{HRE} courses, we further find that \ac{HRE} content only makes up a small fraction of the lectures.

\subsection{Common Threat Models: Novel threats and attribution are overlooked}
\label{sec:results:threatmodels}

The ten hardware security courses without a strong focus on \ac{HRE} address the following prevalent threat models:
\begin{description}
    \item[Cryptographic Secret Extraction.] Using side channels, an attacker with physical access to a device wants to extract a secret value that is being processed or stored on the device. Students may be in the role of either the attacker or the hardware designer implementing side-channel countermeasures. 
    \item[Hardware Trojans.] 
    A malicious modification changes a circuit's functionality to the advantage of the attacker. Students put themselves in the shoes of the attacker or the \ac{OEM} at the end of the supply chain 
    who must detect tampered \acp{IC} before they are deployed. 
    \item[\acl{IP} Infringement.] An unauthorized party discloses, \ie, \enquote{pirates}, the design files or manufactures and sells counterfeits to a third party. 
    Students can play two different defensive roles: either role similar to that of the Trojan case, \textit{detecting} counterfeit \acp{IC}, or the role of a hardware designer aiming to prevent the \textit{production} of counterfeits by incorporating protection mechanisms into the design.
\end{description}

We find that within these threat models, the courses focus on design-time countermeasures and non-destructive analysis looking for known signatures of attacks.
They can therefore be complemented well with \ac{HRE}-specific courses, as \ac{HRE} is a powerful tool for detecting new or particularly subtle manipulations that would otherwise evade detection.
In addition, when tampering is found, \ac{HRE} allows for a deeper understanding of how and where the tamper occurred -- \eg, at the design, manufacturing, or packaging stage -- enabling the removal of bad actors from the supply chain.

\subsection{Educational Approaches}
\label{sec:results:educational}

Having covered \textit{which} topics were conveyed in the courses, we will now consider \textit{how} the course content was delivered.

\subsubsection{Course structure: Combination of lectures with applied projects}
\label{sec:results:educational:structure}

All courses in our corpus have a strong focus on hands-on elements, and all except \coursekey{handson} pair them with a lecture part.
The specifics of the hands-on part vary from course to course:
Many include multiple graded assignments (\coursekey{advtrust}, \coursekey{rub}, \coursekey{cad}), a longer final project (\coursekey{embedded}, \coursekey{introtrust}), or both.
Another common feature are regular in-person lab sessions (\coursekey{introtrust}, \coursekey{zonenberg}, \coursekey{rub}, \coursekey{handson}, \coursekey{carpenter}, \coursekey{phiks}).
In \coursekey{phiks} and \coursekey{zonenberg}, students have access to an extensive hardware inspection laboratory.
However, the two courses differ in their approach to interactivity: While \coursekey{phiks} allows students to use the equipment on their own, the lab sessions of \coursekey{zonenberg} are designated as \enquote{demonstration}.
This suggests that the instructor performed the experiment while the students observed.

\subsubsection{Materials: Wide variation in type and cost}
\label{sec:results:educational:materials}
While several courses use primarily digital environments for their labs and assignments (\coursekey{embedded}, \coursekey{halak}, \coursekey{rub}), others use commercial devices (\coursekey{crypteng}, \coursekey{practical}), and \coursekey{handson} utilizes custom hardware developed for the course.

Two publications stood out in our analysis:
Karam~\etal~\cite{karam2022improving} make a strong case for \enquote{accessible} hardware security training and designed the labs for \coursekey{practical} around low-cost hardware which students can take home, supported by open-source software.
True~\etal~\cite{true2023physical}, in stark contrast, showcase the extensive selection of high-cost on-site laboratory equipment used in \coursekey{phiks}.
Replicating this type of course format at other institutions would require extensive lead time, skilled laboratory equipment operators, and may be prohibitively costly for many institutions.

Hardware simulations are commonly used, especially when physical experiments would be too costly.
For instance, many of the courses in our analysis cover the attacker's perspective of hardware Trojans, \ie, Trojan insertion (\coursekey{trustable}, \coursekey{carpenter}, \coursekey{practical}).
Naturally, none of them go as far as physically manufacturing a trojanized \ac{IC}. 

Two courses (\coursekey{handson}, \coursekey{rub}) have found a middle ground that allows students to tangibly experience the impact of design manipulation:
By deploying a manipulated design on reconfigurable hardware, students can easily test and iterate their implementation.

\subsection{Evaluations of Course Effectiveness}
\label{sec:results:evaluation}

\bgroup
\setlength\tabcolsep{3pt}
\def\arraystretch{.9}
\begin{table}[b]
    \centering
    \caption{Evaluation data presented for each course, using the data type (DT) taxonomy by Švábenský~\etal~\cite{svabensky2020what}. Only courses with an accompanying publication are included.}
    \label{tab:results:evaluation:matrix}
    \begin{tabular}{l*{8}{c}*{2}{>{\centering}b{20pt}}*{4}{c}}
                              &&                    && \multicolumn{4}{c}{\parbox{60pt}{\centering\textbf{Questionnaire (DT1)}}}                                             && \multicolumn{2}{c}{\parbox{40pt}{\centering\textbf{Test Data (DT2)}}} && \textbf{DT3}            && \textbf{DT6}\\
                                                       \cline{5-8}                                                                                         \cline{10-11}  \cline{13-13} \cline{15-15}
        \rot{\textbf{Course}} && \rot{\parbox{1.8cm}{Demo-\\graphics}} && \rot{Pre-survey} & \rot{Post-survey} & \rot{\parbox{1.8cm}{Project-wise Survey}} & \rot{QA Survey} && \rot{Grades}  & \rot{\parbox{1.8cm}{Pre/Post\\Quiz}} && \rot{\parbox{1.8cm}{Student\\Artifacts}} && \rot{Anecdotal}  \\
        \cline{1-1}              \cline{3-3}           \cline{5-8}                                                                                         \cline{10-11} \cline{13-13} \cline{15-15}
        \coursekey{embedded}  &&  -                 && -                & -                 & -                         & -                             && -             & -                            && \fullcirc               && -\rule{0pt}{1em}\\
        \coursekey{halak}     &&  -                 && -                & -                 & -                         & \fullcirc                     && -             & -                            && -                       && \fullcirc        \\
        \coursekey{rub}       && \fullcirc          && \fullcirc        & -                 & \fullcirc                 & -                             && \fullcirc     & -                            && \fullcirc               && -                \\
        \coursekey{handson}   && \fullcirc          &&                  & -                 & -                         & \fullcirc                     && \fullcirc     & -                            && -                       && -                \\
        \coursekey{carpenter} && \fullcirc          && \fullcirc        & \fullcirc         & -                         & -                             && -             & \fullcirc                    && -                       && \fullcirc        \\
        \coursekey{phiks}     &&  -                 && -                & -                 & -                         & -                             && -             & -                            && -                       && \fullcirc        \\
        \coursekey{crypteng}  && \fullcirc          && -                & -                 & -                         & \fullcirc                     && \fullcirc     & -                            && -                       && \fullcirc        \\
        \coursekey{practical} && \fullcirc          && \fullcirc        & \fullcirc         & -                         & -                             && -             & -                            && -                       && -                
    \end{tabular}
    \Description[A matrix mapping courses to evaluation methods]{A matrix with one row for each course and one column for each evaluation method. Dots indicate that a method was used in a specific course, dashes indicate the opposite. The first column indicates the course abbreviation. Column two indicates whether the paper presents demographic information. The remaining eight columns contain the methods. The topics are grouped by data type.}
\end{table}
\egroup

We find no consensus on the extent and rigor of course evaluations, which is consistent with the findings of a 2020 review of publications on the broader field of cybersecurity education~\cite{svabensky2020what}, as well as a recent US-focused review~\cite{crabb2024critical}.
\autoref{tab:results:evaluation:matrix} shows each evaluation method used in the eight courses with an accompanying paper.

The evaluation of \coursekey{phiks} is based entirely on anecdotal evidence.
Other commonly used metrics are student performance as measured by grades or argued from the quality of artifacts submitted, and self-reports of time invested in each project.
\coursekey{rub}, \coursekey{carpenter}, and \coursekey{practical} use student self-reporting in the form of surveys about their level of knowledge (see \cite{olney2023development} for a post-assessment of \coursekey{practical}).

We find the most extensive evaluation in our corpus in \coursekey{carpenter}, where self-reports are triangulated by a quiz accompanying each survey.
However, the publication does not report any statistical tests for significance, and little is known about the design of the quiz items.

The respective evaluations generally find that students rate the courses highly on \ac{QA} surveys (\coursekey{embedded}, \coursekey{halak}, \coursekey{handson}), with students indicating that they learned a lot, but that the workload is high (\coursekey{handson}).
They also note that the majority of students is able to achieve high grades (\coursekey{rub}, \coursekey{handson}, \coursekey{crypteng}) and performs well in projects (\coursekey{embedded}).
However, the grade distribution is sometimes skewed, with a few individual students receiving very low grades (\coursekey{embedded}).

\section{Discussion and Recommendations}
\label{chap:discussion}

Below, we discuss our findings on the content, pedagogical features, effectiveness, and assessment methods of existing hardware security and \ac{HRE} courses based on the research questions outlined in \autoref{par:introduction:rqs}.
From these findings, we derive initial recommendations for efficiently implementing \ac{HRE} training programs.

\subsection{RQ1: Contents and Pedagogical Features}
\label{sec:results:rq1}

\paragraph{Courses focus on non- or semi-invasive attacks and defenses, less on \texorpdfstring{\acs{HRE}}{HRE}}
\label{sec:results:rq1findinga}

Few hardware security courses include (invasive) \ac{HRE} as a core topic, despite its inclusion in the current cybersecurity curriculum guidelines~\cite{jtfcc2018cybersecurity}.
Most focus on non- or semi-invasive attacks and defenses, primarily \acf{SCA}, hardware Trojans, \ac{IP} protection and counterfeits, and fault attacks.
They further convey theoretical knowledge of both \ac{IC} manufacturing and cybersecurity fundamentals.

The three courses primarily covering \ac{HRE} do so with varying focus;
two emphasize physical aspects, such as \acf{PCB} reverse engineering, destructive microscope imaging of \acfp{IC}, and subsequent image processing;
the third primarily covers the analysis of logic circuits, or netlists.

\paragraph{Courses combine lectures with hands-on projects using a variety of materials}
\label{sec:results:rq1b:finding}

All 13 courses place a strong emphasis on practical experience, whether through projects, in-person lab sessions, or technical demonstrations by an instructor.
Hands-on assignments are commonly used to evaluate student performance and account for a large portion of the final grade.
Almost all the courses further combine the practical projects with a lecture series.

The materials chosen to engage students vary widely, as does the cost of those materials.
Some courses primarily use digital environments including hardware simulations.
Others use off-the-shelf or custom hardware, and commercially available laboratory equipment; the latter two requiring the most financial and human resources.
Several instructors make a deliberate choice to use low-cost, portable hardware to increase student access to hardware security education by lowering the financial entry barrier.

\paragraph{Recommendations:}
\label{sec:results:rq1:recommendationa}
We call for action to develop \ac{HRE} courses for inclusion in the curricula of existing cybersecurity programs.
This will require empirically assessing the demand of various stakeholders such as industry, politics, society, and academia.
Subsequently, curriculum guidelines should be enriched with concrete suggestions for effective course packaging for the various topics in \ac{HRE}.

Based on the findings of this review, we offer the following initial suggestions:
Institutions should start by using digital environments and simulated hardware for cost-effective lab sessions.
As budgets allow, incorporating commercially available hardware can enhance hands-on learning, covering concepts such as \ac{PCB} reverse engineering or circuit manipulation based on \acp{FPGA}.
Project design should vary over the length of the course:
Beginner projects should be scaffolded~\cite{collins1988cognitive, reiser2014scaffolding} to guide students step-by-step through novel problems, while later projects should provide less guidance, simulating real-world \ac{HRE} settings. 
This approach allows students to develop the competencies needed to perform \ac{HRE} independently, such as navigating obscure documentation and a high tolerance for frustration.

Instructors of \ac{HRE} training programs that primarily use digital environments and simulations should consider making their course contents available online.
Given the limited number of existing \ac{HRE} courses, this would be a valuable step in rapidly expanding access to a larger student body.

\subsection{RQ2: Evaluating Course Effectiveness and Assessment Methods}
\label{sec:results:rq2}

\paragraph{Limited available evaluation suggests effectiveness, but high workload may hamper learning}
\label{sec:results:rq2:findinga}

Evaluation results indicate that the courses are effective in imparting hardware security and \ac{HRE} knowledge and skills.
The combination of theoretical lectures with practical exercises and projects receives positive feedback in formal \acf{QA} surveys.
Students report substantial learning gains, suggesting the efficacy of this instructional approach.
Grading data show that students generally achieve good or excellent final grades, indicating successful project completion.
However, concerns arise regarding the perceived high workload.
This workload could impose excessive cognitive load and stress on students, potentially compromising overall learning outcomes.

\paragraph{Comprehensive evaluation requires methods beyond student self-assessment}
\label{sec:results:rq2:findingb}

Most hardware security courses are not thoroughly evaluated for learning outcomes beyond relying on student self-reporting.
Such student self-assessments appear to have limited reliability:
Olney~\etal~\cite{olney2023development} found little statistical significance but an overall increase in self-rated knowledge levels across all items, despite some topics not being covered in the course.
This increase may have been influenced by concurrent courses covering related topics and the tendency of students to rate their knowledge higher post-course, reflecting expected learning outcomes and increased confidence after completing the semester.

\paragraph{Recommendations:}
\label{sec:results:rq2:recommendationa}

To maintain a manageable student workload, we recommend offering advanced classes on the \ac{HRE} topics from \autoref{tab:results:coverage:matrix}.
Additionally, widely accessible introductory courses are essential to attract and prepare new students entering the field.

Given the inconsistencies in course evaluation, we propose that educational guidelines for \ac{HRE} should include evaluation criteria to facilitate designing new courses based on concrete evidence of effectiveness.
Testing-based assessments should be used to measure student knowledge beyond self-assessment and participation in formal \ac{QA} surveys should be prioritized.
We advocate for transparently reporting these evaluation results in course-related publications.

\subsection{Limitations}
\label{sec:discussion:limitations}

We acknowledge several limitations of this literature review.
While starting our search for courses from scientific publications ensures replicability and provides insight into their pedagogical considerations, future work should consider additional relevant courses which may be discoverable through university catalogs.
Additionally, we only analyzed publicly available course materials.
Requesting materials from individual instructors could enable more precise insight into course content and pedagogical features.
This analysis further focused on English-language publications only.
Considering that the three dedicated \ac{HRE} courses identified are located in the United States and Germany, the inclusion of works in other languages may help to mitigate cultural bias.
\section{Conclusion and Outlook}
\label{chap:conclusion}

\acf{HRE} is crucial for protecting the global semiconductor supply chain, yet guidance for teaching \ac{HRE} is limited.
Based on our systematic literature review of existing courses, we proposed several recommendations for designing \ac{HRE} training programs that are viable under time and budget constraints, and for evaluating their effectiveness.
Further research is needed to formalize these recommendations for inclusion in cybersecurity curriculum guidelines.
To ensure the continuity of existing programs, we advocate for the integration of hardware security and \ac{HRE} training into mandatory curricula.

Several related concerns warrant further exploration by practitioners and researchers:
Improving student access to \ac{HRE} training requires better discoverability of existing courses and public availability of course materials.
We were able to obtain course materials for only four of the 13 courses, and discovering courses without accompanying publications was difficult.
Additionally, accessible materials often disappear without proper archival when courses are discontinued.
Closer collaboration between academic institutions and industry partners may allow \ac{HRE} materials to be shared more freely, overcoming the \ac{IP} concerns and resulting secrecy that are common in the semiconductor industry.
Establishing a more transparent and coordinated archival system can then make materials more discoverable and accessible, allowing instructors to reuse well-evaluated course materials.
We believe that these steps can make \ac{HRE} education available to a larger student body and will help close the semiconductor workforce gap.

\begin{acks}
    We would like to thank our anonymous reviewers for their constructive feedback on this work. We are also grateful to Sarah Naqvi who was a great help in data analysis.
    This work was supported by the \grantsponsor{sechuman}{PhD School \enquote{SecHuman -- Security for Humans in Cyberspace} by the federal state of NRW, Germany}{https://sechuman.ruhr-uni-bochum.de/}, the \grantsponsor{dfg}{German Research Foundation (DFG)}{https://www.dfg.de/en} within the framework of the Excellence Strategy of the Federal Government and the States - \grantnum{dfg}{EXC 2092 CASA - 390781972}, and the \grantsponsor{rctrust}{Research Center Trustworthy Data Science and Security}{https://rc-trust.ai} (\url{https://rc-trust.ai}), one of the Research Alliance Centers within the UA Ruhr (\url{https://uaruhr.de}).
\end{acks}

\bibliographystyle{ACM-Reference-Format}
\pagebreak
\balance
\bibliography{main}


\end{document}